\documentclass[a4paper]{jpconf}
\usepackage{graphicx}
\usepackage{amsmath}
\usepackage{iopams}
\usepackage{float}
\usepackage[font={small},labelfont={bf}]{caption}

\def\camII@III{II/III}
\def\camIV{IVC$\beta$}
\def\camVI{VI}
\def\camLGN{LGN}
\def\mV{\ensuremath{~\textnormal{mV}}}
\newcommand*{\avgleft}{\left\langle} 
\newcommand*{\avgright}{\right\rangle} 
\newcommand*{\average}[1]{\avgleft#1\avgright}
\def\virgula{\ \textnormal{,}} 

\begin{document}

\title{Information processing occurs via critical avalanches in a model of the primary visual cortex}

\author{G.~S.~Bortolotto$^1$, M.~Girardi-Schappo$^1$, J.~J.~Gonsalves$^1$, L.~T.~Pinto$^2$, M.~H.~R.~Tragtenberg$^1$}
\address{$^1$ Departamento de F{\'i}sica, Universidade Federal de Santa Catarina, 88040-900, Florian{\'o}polis, Santa Catarina, Brazil}
\address{$^2$ Departamento de Engenharia Qu{\'i}mica e de Alimentos, Universidade Federal de Santa Catarina, 88040-900, Florian{\'o}polis, Santa Catarina, Brazil}
\ead{girardi.s@gmail.com}
\ead{marcelotragtenberg@gmail.com}
\address{$ $}
\address{doi:10.1088/1742-6596/686/1/012008}

\begin{abstract}
We study a new biologically motivated model for the Macaque monkey primary visual cortex which presents power-law avalanches after a visual stimulus. 
The signal propagates through all the layers of the model via avalanches that depend on network structure and synaptic parameter.
We identify four different avalanche profiles as a function of the excitatory postsynaptic potential.
The avalanches follow a size-duration scaling relation and present critical exponents that match experiments.
The structure of the network gives rise to a regime of two characteristic spatial scales, one of which
vanishes in the thermodynamic limit.
\end{abstract}

\section{Introduction}
\label{sec:Intro}

Brain criticality has gained wide attention in the last years~\cite{chialvoReview,criticalityBook2014,beggsEditorial}. 
Several studies have shown how multiple cognition features are improved when considering the brain as a critical system.
Some of these features are the optimization of response dynamic range~\cite{kinouchiCopelli,plenzDynRange} and the enhancement of
memory and learning processes~\cite{socPlasticity,scarpettaMem2013}, computational power~\cite{plenzBenefits,arcangelisLearn2010},
information processing flexibility~\cite{mosqueiro2013} and network processing time~\cite{girardiV12015}.

The visual system has such a collection of experimental results~\cite{albrightNeuroRev2000,cliffordVisAdapt2007}
that makes it suitable for validating models. Recently, some authors probed cortical tissues and sometimes specifically
the visual cortex
looking for neuronal avalanches both \textit{in vitro}
and \textit{in vivo}~\cite{beggsPlenz2003,ribeiroCopelli,plenzAvalV12010,shewV1Aval2015}.
We developed a model of the primary visual cortex (V1) based on experimental constraints in order to study
how do the microscopic details of the system shape the avalanche dynamics and statistics.

The stimulation of the retina of our model gives rise to avalanches.
The avalanches spread radially inside each layer and through a branching process across different columns.
The columnar structure and the whole network excitation are evident in the avalanches size and duration distributions which present two distinct
power-law (PL) behaviors for a given \textit{excitatory postsynaptic potential} (EPSP) range.
We computed the exponents of these distributions and checked that they obey Sethna's scaling law~\cite{sethna2001} even
when the system is adjusted outside of the critical regime.

Next section is dedicated to describe the model dynamics and structure. Section~\ref{sec:results} brings results
and discussions concerning the information processing of the network via avalanche activity. After that, we
conclude reviewing the main results and we point some possible extensions for this work.

\section{Model}
\label{sec:model}

\begin{figure}[t!]
\begin{center}
	\includegraphics[width=\textwidth]{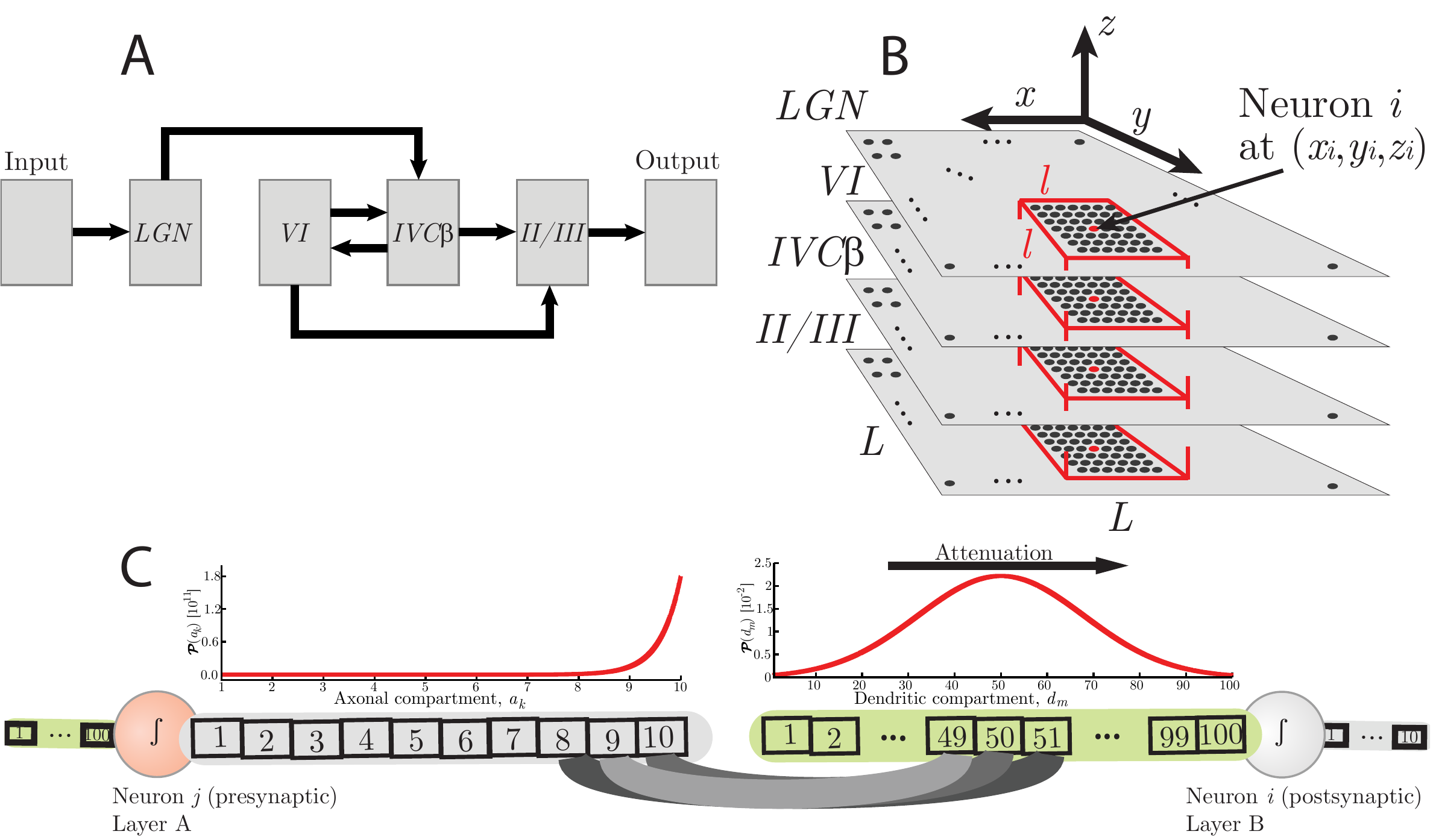}
	\caption{\label{fig:model}The V1 model. 
	A:~Architecture of the network. 
	B:~Spatial organization of the network of $N = 4L^{2}$ neurons. Each neuron in the network is at the center of a column of size $N_{c} = 4l^{2} = 196$
neurons defined by the algorithm used to create synapses. This columnar structure in highlighted in red in Panel B.
   C:~ Compartmental scheme of neurons. Neuronal synapses start preferentially from the end of the axon (left),
as given by the probability $\mathcal{P}(a_{k})$ of choosing a presynaptic axonal compartment $a_{k}$. The dendritic postsynaptic compartment, $d_{m}$, is chosen with a Gaussian probability $\mathcal{P}(d_{m})$ with mean 50 and standard deviation 10.
    Most synapses lay in the middle of the dendrite (right).}
\end{center}
\end{figure}

The model developed by Andreazza \& Pinto  is composed of six square layers~\cite{andreazzaSim2006}.
 The layers are connected to each other following a feed-forward mechanism with a single loop presented in Fig.~\ref{fig:model}A.
No lateral connections are present inside the layers.
The signal propagates from the 
 retina (Input layer) to the secondary visual cortex V2 (Output layer).

The four internal layers have linear size $L$ and correspond to the form recognition pathway within V1
(composed of layers \camII@III, \camIV\ and \camVI)
plus the lateral geniculate nucleus (\camLGN).
The \camLGN\ layer consists of only its parvocellular neurons. The synapses of such neurons are mostly connected to
V1 layer \camIV~\cite{okuskyV11982,lundV11984,callawayV1Review1998,yabutaV11998}.

\subsection{Network structure}

The network has a total of $N_{full}=2N_{io}+N$ elements. The four internal layers
 (\camLGN, \camVI, \camIV\ and \camII@III) are composed by $L^2$ neurons each, 
summing up to a total of $N=4L^2$ neurons.
The neurons are composed of compartmental dendrites and axons connected by a soma
compartment as seen in Fig.~\ref{fig:model}C. Each of the input/output layers has a total of $N_{io}=(10L)^2$ elements.

The Input layer represents the photoreceptors of the retina.
The Output layer consists of axonal terminals which connect to dendrites in V2.
Each neuron of the \camLGN\ is placed in front of a matrix of 100
photoreceptors of the Input layer. A similar structure is present on the other end of the network
where each neuron in layer \camII@III\ randomly sends 100 synapses towards a $10\times10$ matrix in V2.

Layers are positioned parallel to the $x-y$ plane and are stacked along the $z$ direction (see Fig.~\ref{fig:model}B). Each presynaptic neuron $j$ has a spatial position $\vec{r}_j=\left(x_j,y_j,z_j\right)$. The number of synapses that each neuron of each layer sends towards its adjacent layers are fixed according to Table~\ref{tab:synPerPreSyn}.

\begin{table}[t!]
\caption{The quantity of attempted synapses per presynaptic element from a specific layer (rows) to another layer (columns). Presynaptic element may be a photoreceptor (from Input layer) or a neuron (from internal layers). These values are predefined in the beginning of the simulation, but are not exactly achieved in practice because of free boundary conditions.}
\centering
\begin{tabular}{cccccc}
\br
\textbf{From~\textbackslash~To}    & \textbf{\camLGN} & \textbf{\camVI}  & \textbf{\camIV} & \textbf{\camII@III} & \textbf{Output}\\\mr
\textbf{Input}      & 1 & -    & -    & -   & -   \\
\textbf{\camLGN}    & - & -    & 500  & -   & -   \\
\textbf{\camVI}     & - & -    & 1100 & 350 & -   \\
\textbf{\camIV}     & - & 600  & -    & 700 & -   \\
\textbf{\camII@III} & - & -    & -    & -   & 100 \\\br
\end{tabular}
\label{tab:synPerPreSyn}
\end{table}

Every postsynaptic neuron is chosen in the adjacent layer with the use of a bidimensional Gaussian distribution,
$\mathcal{P}_G(x,y;x_j,y_j,\sigma_c)$ 
centered in the presynaptic neuron $(x_{j},y_{j})$ position with standard deviation $\sigma_c=3$ inside a limited region of $l^2=7\times7$ neurons. This mechanism creates a columnar structure  as shown in Fig.~\ref{fig:model}B.

Each column has approximately $N_c=4l^2=196$ neurons.
When taking into account the loop between layers \camIV\ and \camVI, the columns can be
considered to have  $N_c=5l^2=245$ neurons. Such structure is very important for the simulation results.
Synapses sent by neurons on the border of each layer that fall outside of the adjacent layer
are simply ignored (free boundary conditions). 

Once postsynaptic neurons are chosen for every presynaptic neuron, we pick one axonal
compartment $a_k^{(j)}$ of the presynaptic neuron~$j$ with exponential probability
$\mathcal{P}_E(k)=(10/4)\exp\left(10k/4\right)$ and one dendritic compartment $d_m^{(i)}$ 
of the postsynaptic cell~$i$, chosen with Gaussian probability
$\mathcal{P}_G(m;d^{(c)},\sigma_d)$ to form each synapse.
Here $d^{(c)}=50$ and $\sigma_d=10$. These distributions are plotted in Fig.~\ref{fig:model}C.

Synaptic distribution over dendritic compartments is chosen such that the probability of receiving a synapse
in the dendrite is maximum at its center~\cite{williamsV12002} and that any consecutive synapses are uncorrelated.
Axonal compartments distribution is chosen such that most of the synapses come out of the axon's end,
making the signal travel as far as possible.

\subsection{Neuronal and synaptic dynamics}

Each neuron $i$ is composed of a compartmental dendrite 
[$d_m^{(i)}(t);~m=1,\cdots,100$,
Eq.~\eqref{eq:compDendritoV1}], the soma [$v_i(t)$, Eq.~\eqref{eq:compSomaV1}], and
a compartmental axon [$a_k^{(i)}(t);~k=1,\cdots,10$,
Eq.~\eqref{eq:compAxonioV1}].
The action potential advances one compartment per time step $t$. 
This potential comes from the dendrites, passes through the soma and
finally reaches the last axonal compartment:

\begin{equation}
\label{eq:compDendritoV1}
\begin{array}{ll}
    d_1^{(i)}(t+1)&=\lambda E\sum\limits_{j,n}{a_n^{(j)}(t)}\virgula\\
    d_k^{(i)}(t+1)&=\lambda\left[ d_{k-1}^{(i)}(t)+E\sum\limits_{j,n}{a_n^{(j)}(t)}\right],\textnormal{for }k>1,
\end{array}
\end{equation}
\begin{equation}
\label{eq:compSomaV1}
v_i(t+1)=\left\{
\begin{array}{ll}
    \Theta\left(d_{100}^{(i)}(t)-v_T\right) & \textnormal{, if }v_i(t)=0\virgula\\
    -R & \textnormal{, if }v_i(t)=1\virgula\\
    v_i(t)+1 & \textnormal{, if }v_i(t)<0\virgula
\end{array}
\right.
\end{equation}
\begin{equation}
\label{eq:compAxonioV1}
\begin{array}{ll}
    a_1^{(i)}(t+1)&=\Theta\left(v_i(t)\right)\virgula \\
    a_k^{(i)}(t+1)&=a_{k-1}^{(i)}(t), \textnormal{for }k>1,
\end{array}
\end{equation}
where
$\lambda=0.996$ is the attenuation constant
(following experimental results~\cite{williamsV12002}),
$v_T=10\mV$ is the firing threshold,
$R$ is the refractory period (which is sufficient to avoid self-sustained activity within the interlayer loop)
and $E>0$ is the EPSP (of the order of~$1\mV$~\cite{williamsV12002,songEPSP2005,lefortEPSP2009}). 
The double sum in Eq.~\eqref{eq:compDendritoV1} is over all the axonal compartments of presynaptic neuron $j$ connected
to the dendritic compartment of the postsynaptic neuron.

The different amount of dendritic and axonal compartments for each neuron takes into account
the different velocities of signal propagation for a time step of $1~\mu$s.
In addition, the soma time step is of $1$~ms, so that avalanches have a larger time scale than the propagation of the signal
through the neurons compartments.
The travelled distance in the dendrites is about $100~\mu$m and for the axons
$1000~\mu$m~\cite{deutschNervous1993}. Initial conditions are
$a_k^{(i)}(0)=d_m^{(i)}(0)=v_i(0)=0\ \forall\ (k,m,i)$ for all the neurons of the internal layers.
A square of $30\times30$ photoreceptors is flashed in the first time step in order to spark activity.
Results are stable for different locations of the retina activation stimulus.

\begin{figure}[b!]
\begin{center}
	\includegraphics[width=80mm,height=75mm,keepaspectratio]{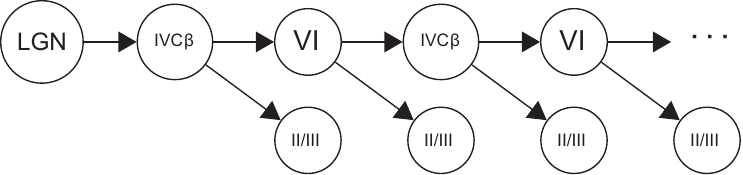}
	\hspace{2pc}%
    \begin{minipage}[b]{14pc}
        \caption{\label{fig:algolayers}Schematic representation of signal branching process in the network. This follows from the
        network architecture in Fig.~\ref{fig:model}.}
    \end{minipage}	
\end{center}
\end{figure}

\section{Results and discussion}
\label{sec:results}

We performed about one hundred simulations for each network linear size, $L$, and each EPSP, $E$.
The activity of the network is the sum of all the neurons firings at each
time step, $A(t)=\sum_{i=1}^{N}{\delta_{v_i(t),1}}$, where $\delta_{a,b}$ is the Kronecker delta.
An avalanche size is defined as the total activity between two consecutive instants of inactivity (similarly
to the experimental procedure~\cite{ribeiroCopelli,priesemannSub2014}),
and the avalanche duration is simply the time interval between these consecutive instants of inactivity,
\begin{align}
s(n+1)&=\sum_{t=t_n}^{t_{n+1}}{A(t)}\ ,\\
T(n+1)&=t_{n+1}-t_n\ ,
\end{align}
where $A(t_n)=A(t_{n+1})=0$. 

On the critical point, the distributions of $s$ and $T$ are assumed to scale as~\cite{pruessnerSOC2012}
\begin{align}
\label{eq:sizeDist}
\mathcal{P}(s)&\sim s^{-\alpha}\mathcal{G}_s\left(s/s_c\right)\ ,\\
\label{eq:timeDist}
\mathcal{P}(T)&\sim T^{-\tau}\mathcal{G}_T\left(T/T_c\right)\ ,
\end{align}
where $s_c$ and $T_c$ are the cutoff values of $s$ and $T$, $\mathcal{G}_{s,T}(x)$ are scaling functions that describe how the cutoff
of the distributions scale with system size and $\alpha$ and $\tau$ are scaling exponents that should follow Sethna scaling
relation~\cite{sethna2001}:
\begin{equation}
\label{eq:sethna}
a=\dfrac{\tau-1}{\alpha-1}\ ,
\end{equation}
where $a$ is a growth exponent relating avalanches size and duration,
\begin{equation}
\label{eq:avalGrowth}
\average{s}\sim T^a\ .
\end{equation}
The scaling law
of Eq.~\eqref{eq:sethna} must be obeyed if Eqs.~\eqref{eq:sizeDist} and~\eqref{eq:timeDist} hold within a
large enough interval of $s$ and $T$.

\begin{figure*}[t!]
    \includegraphics[width=\textwidth]{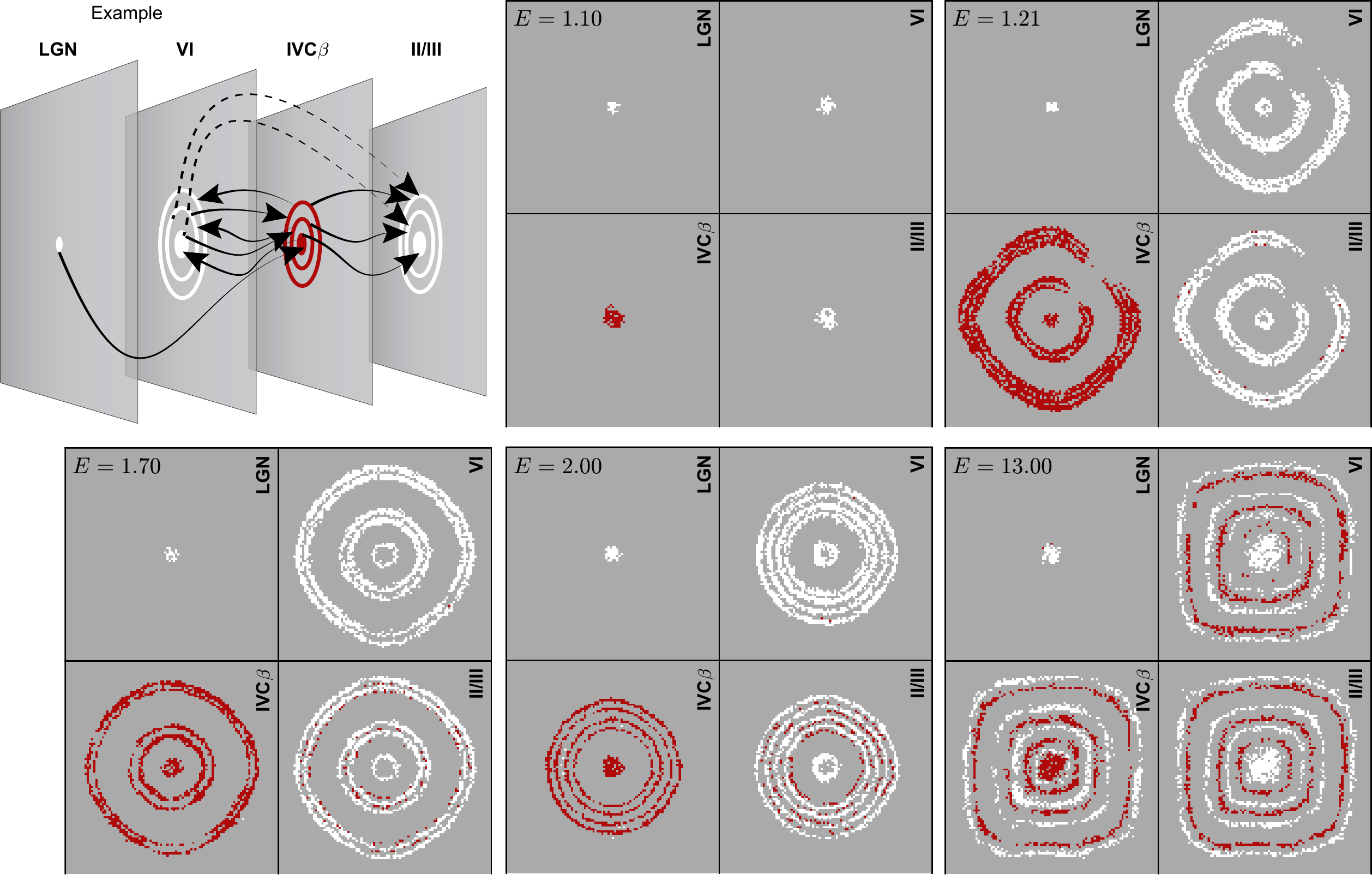}
	\caption{\label{fig:netact}Spatial profile of avalanches in the internal layers for different $E$.
    Each circle is a snapshot of the network in a different time step.
    Top left panel illustrates the temporal order of events which is useful to analyze the other panels:
    Activity starts in the center of the \camLGN\ and then follows the
    directions pointed by the arrows. Circle colors are used to indicate temporal sequence: activity in red always follows activity
    in white and so on, although not in the immediate next time step.
	$E=1.10\mV$: the signal travels through all layers but is unable to reach the border of the layers; layer \camII@III\ spikes
    only together with layer \camVI. In fact, this means that avalanches are spreading only inside columns
    in a sort of branching process (see Fig.~\ref{fig:algolayers}).
    Starting from $E=1.12\mV$, activity has probability to reach the border of the layers.
	For $E=1.21\mV$, the activity always reaches the border;
    layer \camII@III\ spikes together with \camVI, although some of \camII@III\ neurons
    spike together with \camIV\ due the signal coming from \camVI.
	The panels with $E=1.70\mV,2.00\mV$ show how the entanglement of activity in layers \camIV\ and \camII@III\ strengthens
    for larger EPSP; the avalanches become larger due to less inactivity intervals (see Fig.~\ref{fig:timeseries_aval}A).
	For $E=13.00\mV$ activity in layer \camVI, \camIV\ and \camII@III are so entangled that a single large avalanche
    appears and radially spreads inside the network.}
\end{figure*}

\subsection{Spatio-temporal profile of avalanches}
\label{sec:Net}

Activity is initiated in the \camLGN\ and sent towards layer \camIV\ inside a small localized region due to the
columnar structure of the network. This small activated region in layer \camIV\ acts as a seed for firing localized activity
in the adjacent layers, \camVI\ and \camII@III. The activity of layer \camII@III\ is just sent over to V2 while
activity from layer \camVI\ serves as another seed for firing localized activity again in layer \camIV.
Now, the neurons that have already fired in layer \camIV\ are refractory, so only the neurons around the first seed region
will fire this time. This process constitutes a branching process between layers, as illustrated in Fig.~\ref{fig:algolayers}
and in Fig.~\ref{fig:netact} top left panel.

As a consequence of the branching process, the activity spreads as spiking circular waves inside each layer.
The spatial profile of these circular avalanches is in Fig.~\ref{fig:netact}.
The top left panel illustrates how the activity spreads throughout the network internal layers starting in the \camLGN.
Red and white colors indicate the sequence of time steps, although not necessarily immediate consecutive time steps.
Activities represented in red always come after activities represented in white and \textit{vice-versa}.

Two features should be noticed with the help of Fig.~\ref{fig:netact}:
firstly it shows that there is a qualitative change in the behavior of the network as $E$ increases:
for $E<1.12\mV$ the signal does not reach the border of the layers.
For $1.12\leq E\leq1.19\mV$, there is a vanishing probability of reaching the borders whereas for $E>1.19\mV$ the activity
always reaches the border. 
This behavior indicates that $E=1.19\mV$ is a critical point.

\begin{figure}[t!]
\begin{center}
\includegraphics[width=\textwidth]{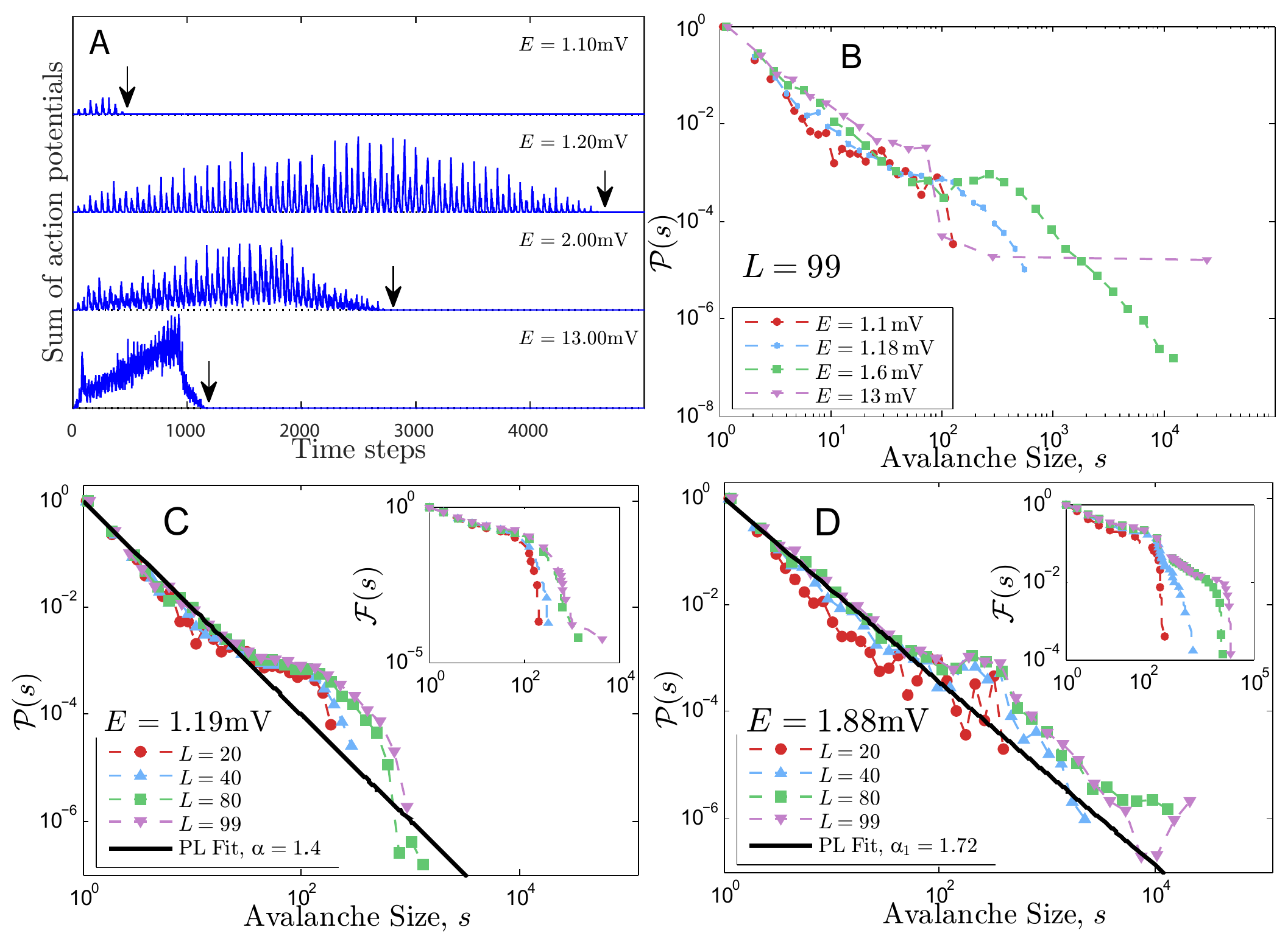}
	\caption{\label{fig:timeseries_aval}A:~Temporal profile of the avalanches for different $E$. Arrows mark
    total processing time;
	B:~Distributions $\mathcal{P}(s)$ of avalanches sizes for four typical $E$ values, one in each regime
    described in the text. An avalanche size is the total activity between two consecutive instants of inactivity in Panel A;
	C:~$\mathcal{P}(s)$ for multiple network sizes $L$ on the critical point $E=1.19\mV$
    (notice how the cutoff scales with system size);
	D:~$\mathcal{P}(s)$ for different $L$ on the weakly ordered regime for $E=1.88\mV$
    (notice the bump in the middle of the distribution localized around $s=N_c\approx200$ neurons);
	The insets in panels C and D show the corresponding complementary cumulative distributions $\mathcal{F}(s)$
    for each $L$ and $E$. Dashed lines are only present to guide the eyes.}
\end{center}
\end{figure}

Secondly, notice that as $E$ increases, layer \camII@III\ activity starts to happen simultaneously with activity in both
layers \camVI\ and \camIV\ because less and less presynaptic neurons (in the seed region)
are needed to cause any neuron to fire. Such profile reduces the silence intervals between consecutive
avalanches (see Fig.~\ref{fig:timeseries_aval}A).
Thus for large $E=13\mV$, the activity spatial profile reduces to a two-dimensional
radial wave front propagating simultaneously in the three layers via a single very large avalanche.

Columns therefore play a significant role by propagating activity only for $E\geq1.12\mV$ but yet with $E$ sufficiently small so that
a neuron needs a considerable amount of presynaptic active neurons to pass on the signal forward. 

The temporal profile of avalanches is shown in Fig.~\ref{fig:timeseries_aval}A for four typical EPSP.
Notice that there is also a qualitative change in the activity temporal profile as $E$ increases.
The network processing time (the time interval between stimulation and the last spike in the network)
continuously increases for increasing $E<1.21\mV$. However, the variance of the processing time is maximal
for the whole interval $1.12\leq E\leq1.19\mV$. The maximum variance for a finite interval indicates that the system
is very flexible to process information inside this region and also might signalize the presence of a critical Griffiths
phase~\cite{griffiths1969,vojtaGrifRev2006}.

Processing time is maximum for $E=1.21\mV$
and decreases asymptotically for increasing $E>1.2\mV$ because the activity across layers becomes entangled.
The entanglement of activity decreases the waiting times between avalanches. Such behavior is evident for $E=2.0\mV$,
in which there is a large avalanche ranging from about $t\approx900$ until $t\approx2100$ time steps,
and also for $E=13\mV$.

\begin{figure*}[t!]
\includegraphics[width=\textwidth]{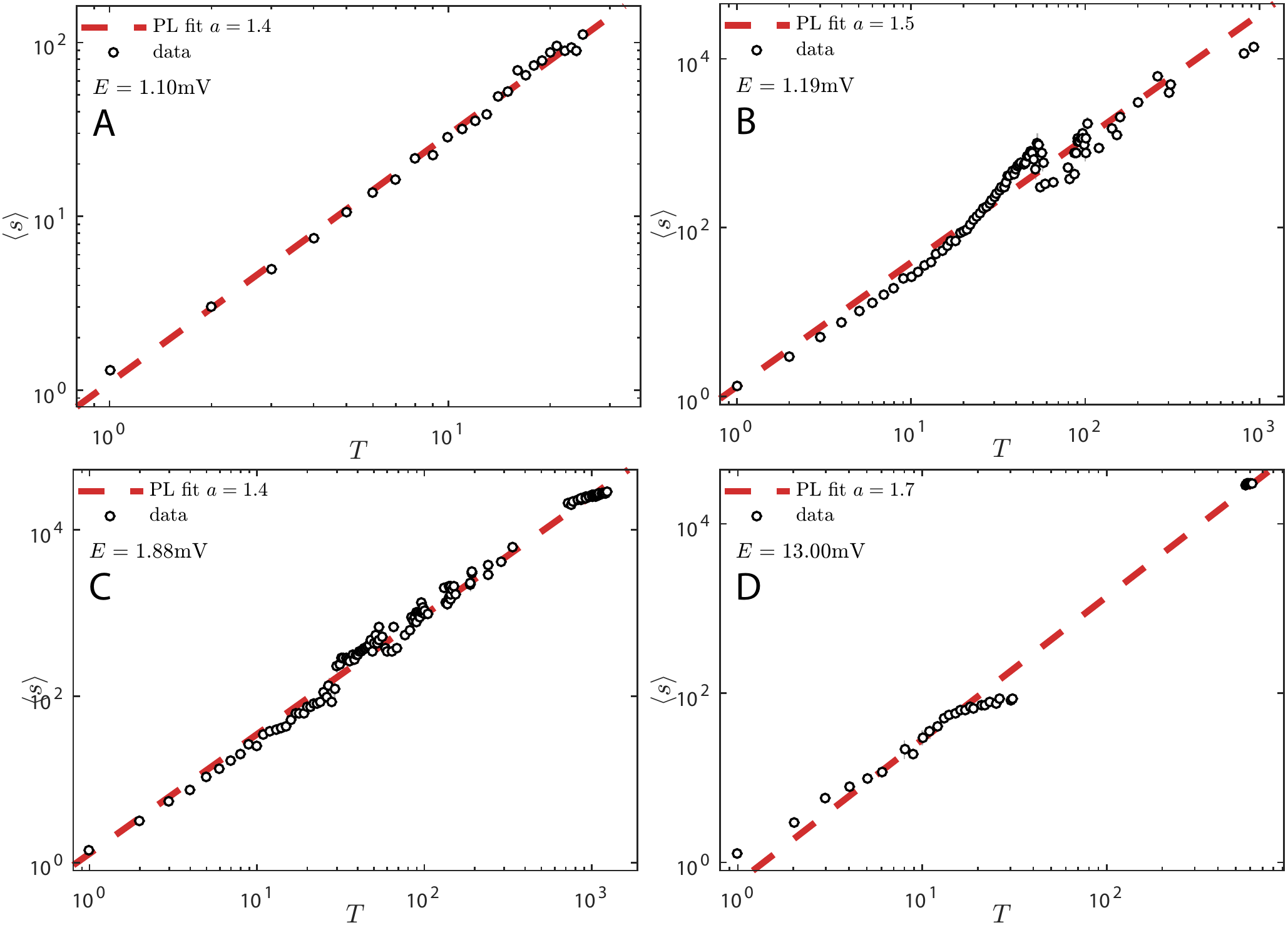}
	\caption{\label{fig:sethna} The size-duration scaling relationship for the avalanches for different $E$.
    The points are the simulation data and the dashed lines are the corresponding fits for $\average{s}\sim T^a$
    yielding $a_{Fit}$ in Table~\ref{tab:sethtable}.
    A:~$E=1.1\mV$, only very small avalanches are present;
    B:~$E=1.19\mV$ is the critical point and presents PL distributed avalanches for all scales;
    C:~$E=1.88\mV$ is in the weakly ordered regime and presents a small gap between very large avalanches and the remaining ones
    (each large avalanche corresponds to a new realization of the simulation);
    D:~for $E=13\mV$ there is a huge gap between the many very large avalanches and some remaining small avalanches,
    where each large avalanche is a simulation realization.}
\end{figure*}

\subsection{Avalanche distributions}
\label{sec:Aval}

We computed the normalized probability and complementary cumulative distributions of the
avalanche sizes and duration. The complementary cumulative distribution (henceforth referred only as 
cumulative distribution) is the probability of measuring an avalanche of size $s$ or greater.
 
In our model the information processing may have four different avalanche profiles~(see Fig.~\ref{fig:timeseries_aval}B).
First, for $E<1.12\mV$ only small avalanches occur. The signal is still able to travel throughout all the layers,
but only a few neurons are in fact activated -- this is the \textit{disordered} regime (or subcritical phase).
Second, the region for $1.12\leq E\leq1.19\mV$ is \textit{critical}: there are many small and large avalanches and each
layer activity is disentangled from the other, as discussed in Section~\ref{sec:Net} and shown in Fig.~\ref{fig:netact}.
Fig~\ref{fig:timeseries_aval}C shows the PL fit of the avalanche size distribution on the critical point.
Such distributions equal all the distributions in the critical $E$ range.
Notice how the largest avalanche, and hence $s_c$ and $T_c$, scale with system size.

The third behavior lies in the range $1.20\leq E\leq2.00\mV$, where large avalanches begin to dominate network activity,
making the distributions tails heavier as $E$ is increased.
Fig~\ref{fig:timeseries_aval}D shows the distribution for $E=1.88\mV$ and many system sizes.
In this EPSP range, we can identify two different PL behaviors, one for small avalanches and another for large ones
due to the entangling of activity across different layers.
The characteristic bump in the middle of the distribution happens exactly for avalanches of size $s=N_c\approx200$.
Thus, small avalanches happen mostly because of disentangled activity inside the network columns 
whereas large avalanches are due to activity that span multiple layers. We call this regime \textit{weakly ordered}.

Finally, for $E > 2.00\mV$ the network is swept by one large avalanche (see Fig.~\ref{fig:timeseries_aval}A and
Fig.~\ref{fig:timeseries_aval}B)
that spans through all the layers simultaneously during the entire simulation time.
The small avalanches regime is still a PL (purple triangles in Fig~\ref{fig:timeseries_aval}B is a PL for $s\lesssim100$),
although these avalanches are just the spark in the \camLGN\ used to start activity.
This behavior is stable for any initial condition, so this is the \textit{strongly ordered} regime.
Together, weakly and strongly ordered regimes make up a supercritical phase.

We calculated the critical exponent $\alpha$ [Eq.~\eqref{eq:sizeDist}] in Figs.~\ref{fig:timeseries_aval} B, C and D
with the collapse of the cumulative
distributions for different system sizes, $L$. This calculation is consistent with a Maximum Likelihood PL fitting
to the PL regime of the distributions~\cite{girardiV12015}.
The same procedure was applied to calculate exponent $\tau$ [Eq.~\eqref{eq:timeDist}].
Some values for $\alpha$ and $\tau$ are presented in Table~\ref{tab:sethtable} (second and third columns).
These values are close to the ones obtained in experiments~\cite{beggsPlenz2003,shewV1Aval2015}
and are close to avalanche exponents of absorbing state phase transitions~\cite{munozExpAbsState1999}.
The collapse of avalanche distributions also yield dynamical spreading exponents
that describe how $s_c$ and $T_c$ increase with $L$. The growth exponent of the largest avalanche
is calculated elsewhere along with a rigorous study of the discussed phase transition~\cite{girardiV12015}.

We find that Sethna's relation holds even outside of the critical regime, contrary to commonly believed~\cite{shewV1Aval2015}.
We calculate Sethna exponent $a$ for the four regimes of $E$ using Eq.~\eqref{eq:sethna} using the values obtained
for $\tau$ and $\alpha$ yielding $a_{Dist}$, the fourth column in Table~\ref{tab:sethtable}.
We also fit the curve Eq.~\eqref{eq:avalGrowth} to data in Fig.~\ref{fig:sethna},
yielding $a_{Fit}$, the fifth column in the same table.
The percentual error between these two values for $a$ is presented in column six.
The errors for $a_{Dist}$ are inherited from the calculation of $\tau$ and $\alpha$.
These scaling exponents are obtained with a fit on the PL regions of the avalanches size and duration distributions,
each having associated errors of $5\%$. Also, the values of $a_{Fit}$ have associated error of $5\%$.
We can then consider the relative errors between the two quantities to be negligible.
The cutoff $s_c$ for both distributions is small, of the order $L^{D'}$, with $D'=1.1$~\cite{girardiV12015},
and this reflects on the fairly high error values.

\begin{table}[t!]
\caption{The duration and size power-law distribution exponents are shown in columns $\tau$ and $\alpha$, respectively.
Column $a_{Dist}$ is obtained with the use of Eq.~\eqref{eq:sethna}. The values of $a_{Fit}$ are obtained by fitting
simulation data in Fig.~\ref{fig:sethna}. The percentual error between these two approaches is shown in the last column.
$\tau$, $\alpha$ and $a_{Fit}$ have an associated error of~$5\%$ each.}
\centering
\label{tab:sethtable}
\begin{tabular}{cccccc}
\br
\textbf{$E$ (mV)}   &\textbf{ $\tau$} &\textbf{ $\alpha$} & \textbf{$a_{Dist}$} & \textbf{$a_{Fit}$} & \textbf{Error (\%)}  \\ 
\mr
\textbf{1.10} & 1.788 & 1.602     & 1.309 & 1.426 & -8.17 \\ 
\textbf{1.19} & 1.596 & 1.390     & 1.531 & 1.470 & 4.19  \\ 
\textbf{1.88} & 1.737 & 1.491     & 1.500 & 1.434 & 4.58  \\ 
\textbf{13.0} & 1.520 & 1.350     & 1.483 & 1.688 & -12.1 \\
\br
\end{tabular}
\end{table}

\section{Conclusion}

We studied a model presenting several dynamical and structural features
of the primary visual cortex. Some of which have been proved to be of fundamental importance
in the avalanche dynamics of the system: the columnar structure and the extended body of the neurons. 
The first is needed for the signal to activate the entire network
even when the excitatory postsynaptic potential is not strong enough to have a one-to-one neuronal activation.
The latter generates the intrinsic avalanche dynamics of the system since the interval between
avalanches is a consequence of the propagation of action potentials in dendritic and axonal compartments.

We identified four regimes of activity in the system: a disordered regime, in which avalanches are not able
to reach system lateral boundaries and thus avalanches also do not scale with system size
(nevertheless small avalanches are PL distributed);
a critical regime, in which avalanches are PL distributed and avalanche cutoffs scale with system size;
a weakly ordered regime, in which avalanches are PL distributed and are processed in two characteristic size and time scales --
one corresponding to the amount of neurons in a column and the other to the system boundary;
and finally a strongly ordered regime, in which there is a single large dominating avalanche which
scales with system size, although the spark avalanches that initiate activity are also PL distributed.

We also verified that Sethna's relation holds for the four regimes and thus we claim that
having PL distributed avalanches (both in size with exponent $\alpha$ and in duration with exponent $\tau$)
that follow the scaling relation $a=(\tau-1)/(\alpha-1)$ is not sufficient to tell whether the system is
in the critical state. Instead, a good criterion is to have PL avalanches with cutoff that scale with system size.
Nevertheless, scaling of the cutoff is only a weak criterion to determine criticality. The strong
criterion would be to define an order parameter and its associated susceptibility
and then check for the divergence of susceptibility~\cite{girardiV12015}.

Our next steps will be to modify some aspects of the model, e.g.
changing the form of the input signal
in order to to see if the behavior and relationships presented by the network are maintained;
adding heterogeneity in $E$ parameter in order to reproduce its real distribution in the cortex~\cite{songEPSP2005,lefortEPSP2009};
adapting the excitatory field of each neuron in deeper layers;
using synaptic dynamics or plasticity~\cite{shewV1Aval2015,levina,ariadneDynSyn2015} to model adaptability~\cite{kohnVisAdapt2007};
and adding lateral inhibition.

Moreover, we hope to provide here a kinematic framework for microscopic cortical modeling.
We conjecture that our model pertains
to Dynamical Percolation-like universality class
because its dynamics resembles that of the generalized
epidemic process with immunization~\cite{noneqPhaseTrans2008,bonachela2},
which is part of a broad universality class of absorbing state phase transitions~\cite{munozExpAbsState1999}.

\ack
We thank the organizers of the VIII BMSP for the opportunity to show some of the results obtained in this work. Authors GSB and MGS
thank the financial support of FAPESC and CNPq agencies, respectively. We also thank S. Boettcher for valuable discussions.

\section*{References}
 \newcommand{\noop}[1]{}
\providecommand{\newblock}{}

\end{document}